\documentclass[]{interact}

\usepackage{epstopdf}
\usepackage[caption=false]{subfig}

\usepackage[numbers,sort&compress]{natbib}
\bibpunct[, ]{[}{]}{,}{n}{,}{,}
\renewcommand\bibfont{\fontsize{10}{12}\selectfont}

\theoremstyle{plain}

\theoremstyle{definition}

\theoremstyle{remark}

\usepackage{float}
\usepackage{booktabs}
\usepackage{graphics}
\graphicspath{ {Fig/} }
\usepackage{listings}
\usepackage{url}
\usepackage{rotating}
\usepackage{array}
\usepackage{colortbl}
\usepackage{enumerate,multicol}
\usepackage{amsmath, amssymb, amsthm}
\usepackage{enumitem,bm}
\usepackage{parcolumns}
\usepackage{graphicx}
\usepackage{multirow}
\usepackage{makecell}
\usepackage{fancyvrb} 
\usepackage{listings,color} 
\usepackage{moreverb}

\renewcommand\abstractname{ABSTRACT}

\begin{document}

\title{Bayesian Hierarchical Models for the Prediction of Volleyball Results}

\author{
\name{Andrea Gabrio\thanks{E-mail: andrea.gabrio.15@ucl.ac.uk}}
\affil{Department of Statistical Science, University College London,~UK}
}

\maketitle

\begin{abstract}
Statistical modelling of sports data has become more and more popular in the recent years and different types of models have been proposed to achieve a variety of objectives: from identifying the key characteristics which lead a team to win or lose to predicting the outcome of a game or the team rankings in national leagues. Although not as popular as football or basketball, volleyball is a team sport with both national and international level competitions in almost every country. However, there is almost no study investigating the prediction of volleyball game outcomes and team rankings in national leagues. We propose a Bayesian hierarchical model for the prediction of the rankings of volleyball national teams, which also allows to estimate the results of each match in the league. We consider two alternative model specifications of different complexity which are validated using data from the women's volleyball Italian Serie A1 2017-2018 season. 
\end{abstract}


\section{Introduction}\label{intro}
A considerable amount of research has focused around the analysis of sport data, mostly for the prediction of sport scores in each game and team rankings in national leagues. Examples of different types of applications of statistical models to sport data include football~\citep{Dixon1997,Lee1997,Karlis2003,Joseph2006,Baio2010,Tsokos2019}, but also basketball~\citep{Magel2013} and hockey\citep{Roith2013}. However, there is a shortage of published works on the analysis of volleyball, and in particular women's volleyball data. Notable exceptions include: \cite{George2008}, who explored the differences in playing characteristics between winning and losing teams in the international federation of men’s beach volleyball world tour tournament; \cite{Gaetan} and \cite{Zhang2016}, who examined the overall performance of volleyball teams based on many game factors using data from the men's Italian serie A and the women's National Collegiate Athletic Association (NCAA) Division I volleyball league, respectively; \cite{Akarccecsme2017} and \cite{Tumer2017}, who used different approaches to either analyse the results from the games or to predict future team rankings in the Turkish men’s and women's volleyball~leagues. 

A potential factor which limited the analysis of volleyball data in the literature is the fact that the playing rules and the scoring system vary across countries and have also changed over the years, e.g.~from the "side-out"  to the "rally" scoring system. According to the current regulations of the Federation International de Volleyball~\citep{FIVB}, volleyball games have six players on each side. Six rotational spots on the court are changed every time a particular side serves the ball. The aim is to deliver the ball over the net and ground it, or the ball touches on the ground of the opposing side, while preventing the ball from touching the ground on their own side. The volleyball game begins with a "serve" of the ball. Then players take turns rotating around the court so each player has a chance to serve the ball. The members of the opposing team will attempt to "save" it from hitting the ground and knock it to one of their own players or hit it back over the net. Play will continue until one team fails by allowing the ball to touch the ground in its own court or they correctly return it to the opposing court. A point is awarded to the serving team if the opposing team makes a mistake. If the serving team fails, however, then the receiving team has control of the ball and becomes the serving team. 

According to the current scoring system, matches are played until a team wins a set of three games, with each game typically going to $25$ points. However, if the two teams won $2$ games each, the third game then goes only up to $15$ points. A team must win each game by two points. If the score is tied with even numbers, both teams have to continue playing the game until a $2$-point lead is obtained. Otherwise, points keep accumulating until one team wins with a margin of victory of two points, even if the score is greater than $25$ or $15$ points. When playing in professional volleyball leagues, the teams get points according to set numbers at the end of all matches in the league. More specifically, the team points are awarded as follows: if the match is won 3-0 or 3-1, 3 points are assigned to the winner and 0 points to the loser; if the match is won 3-2, 2 points are assigned to the winner and 1 point to the~loser. 

In this paper, building on the previous literature on modelling sport data, we propose a novel Bayesian hierarchical modelling framework for the analysis of volleyball data which allows to jointly predict match results and team rankings in national leagues. We use data from the women's volleyball Italian Serie A1 2017-2018 season as a motivating example to implement and validate the proposed framework. The rest of the article is structured as follows: Section~\ref{data} introduces the notation and the data used for the analysis. Section~\ref{modelframe} and Section~\ref{modelcorr} describe the modelling framework and two alternative model specifications implemented for the analysis of the volleyball data. Section~\ref{results} summarises the results in terms of parameter estimation as well as outcome prediction and assesses the relative performance of the models. Finally, Section~\ref{results} presents some issues and possible extensions that can be the material for future work, while Appendix~\ref{A1} includes the \texttt{JAGS} code for our analysis.

\section{Data from the Italian Serie A1~2017-2018}\label{data}
Data were obtained from the Italian women's volleyball federation web site (\url{http://www.legavolleyfemminile.it/}) and include different types of in-game statistics related to each of the $132$ matches of the regular season 2017-2018 for all $12$ teams in the league. Generally, there are six categories of volleyball statistics which have been found to be associated with match results~\citep{George2008}, which can be broadly distinguished in: attack, setting, serving, passing, defense, and blocking. Data on variables related only to four of these categories are included in the analysis since for some of the categories data are not publicly available on the federation website. In particular, for each team and match of the women's Italian Serie A1 regular season 2017-2018, the following set of statistics are included in our analysis (in brackets the names assigned to each variable):

\begin{description}
\item[Match.] The number of the match.
\item[Teams.] The names of the home team ($home \; team$) and away team ($away \; team$) for each match, together with a unique numeric score assigned to each team regardless of whether it is the hosting ($h$) or visiting ($a$) team. 
\item[Points.] Total number of points scored by the home team ($y_h$) and by the away team~($y_a$).
\item[Sets.] Total number of sets won by the home team ($s_h$) and by the away team ($s_a$) for each match, and a binary indicator $d^{s}$ telling whether the $5$-th set was played ($d^{s}=1$) or not ($d^{s}=0$). Additionally, we define a second binary indicator $d^{m}$, which takes value $1$ or $0$ according to whether the home team won or lost the match, respectively. 
\item[Serve.] Total number of serves ($ser^{tot}_t$), total number of perfect serves or "aces", i.e.~ending in a point for the serving team ($ser^{ace}_t$) and total number of serving errors ($ser^{err}_t$), i.e. ending in a point for the opponent team, for $t=(h,a)$. Serve efficiency is then defined as $ser^{eff}_t=\frac{ser^{ace}_t-ser^{err}_t}{ser^{tot}_t}$.
\item[Defence.] Total number of digs ($def^{tot}_t$), total number of perfect "digs", i.e.~defence action causing the ball to follow a trajectory defined according to specific criteria ($def^{per}_t$) and total number of digging errors ($def^{err}_t$). Defence efficiency is then defined as $def^{eff}_t=\frac{def^{per}_t-def^{err}_t}{def^{tot}_t}$.
\item[Attack.] Total number of attacks ($att^{tot}_t$), total number of perfect attacks, i.e.~attacks ending in a point for the attacking team ($att^{per}_t$) and total number of attacking errors ($att^{err}_t$). Attack efficiency is then defined as $att^{eff}_t=\frac{att^{per}_t-att^{err}_t}{att^{tot}_t}$.
\item[Block.] Total number of blocks ($blo^{tot}_t$), total number of perfect blocks, i.e.~blocks ending in a point for the blocking team ($blo^{per}_t$) and total number of blocking errors or "invasions" ($blo^{err}_t$). Block efficiency is then defined as $blo^{eff}_t=\frac{blo^{per}_t-blo^{err}_t}{blo^{tot}_t}$.
\end{description}

Table~\ref{tab1} shows the structure of the dataset, which includes different types of in-game statistics associated with each game (we use the term $\ldots$ to indicate that the data related to some of the matches and variables have been omitted from the table for clarity). 
\begin{table}
\caption{Summary of the type of data used in the analysis of the Italian Serie A1 women's volleyball league 2017-2018.}\label{tab1}
\centering
\scalebox{0.85}{
\begin{tabular}{cccccccccccccc}
\toprule
 $i$ & $home \; team $ & $away \; team $ & $h$ & $a$ & $y_h$ & $y_a$ & $s_h$ & $s_a$ & $d^s$ & $d^m$ & $ser^{eff}_h$ & $ser^{eff}_a$ & $\ldots$\\
\midrule
 1 & Casalmaggiore & Conegliano & 3 & 4 & 60 & 75 & 0 & 3 & 0 & 0 & -0.17 &  -0.06 & $\ldots$\\
 2 & Novara & Busto Arsizio & 8 & 2 & 113 & 108 & 3 & 2 & 1 & 1 & -0.09 &  -0.10 & $\ldots$\\
 3 & Piacenza & Scandicci & 10 & 12 & 40 & 75 & 0 & 3 & 0 & 0 &-0.07 &  0.03 & $\ldots$\\
 4 & San Casciano & Monza & 11 & 7 & 96 & 94 & 3 & 1 & 0 & 1 & -0.07 &  -0.07 & $\ldots$\\
 $\ldots$ &  $\ldots$ &  $\ldots$ &  $\ldots$ &  $\ldots$ &  $\ldots$ &  $\ldots$ &  $\ldots$ &  $\ldots$ &  $\ldots$ & $\ldots$ &  $\ldots$ &   $\ldots$ & $\ldots$\\
 129 & Casalmaggiore & Piacenza & 3 & 10 & 105 & 103 & 2 & 3 & 1 & 0 & -0.04 & -0.03 & $\ldots$\\
 130 & Scandicci & Busto Arsizio & 12 & 2 & 110 & 98 & 3 & 2 & 1 & 1 &-0.06 & -0.06 & $\ldots$\\
 131 & Filottrano & San Casciano & 5 & 11 & 113 & 108 & 3 & 2 & 1 & 1 & -0.08 & -0.03 & $\ldots$\\
 132 & Novara & Legnano & 8 & 6 & 75 & 57 & 3 & 0 & 0 & -0.08 & 1 &-0.07 & $\ldots$\\
\bottomrule
\end{tabular}
}
\end{table}
For example, the first row contains information about the first match of the season between the home team \textit{Casalmaggiore} and the away team \textit{Conegliano}, which are assigned with the unique numeric scores $3$ and $4$, respectively. The match ended with a total point score of $60$ to $75$ in favour of the away team, who won a total of three sets without losing any set. 

In our analysis we also consider the efficiency measures of the different statistics available for each team when playing at home and visiting: serve and attack efficiency ($ser^{eff}_t$ and $att^{eff}_t$) -- typically related to the offensive strength of each team -- and block and defence efficiency ($blo^{eff}_t$ and $def^{eff}_t$) -- typically associated with the defensive strength of each team. These variables are included in the model to improve the estimation of the attack and defence performances of the teams and the prediction of the results for future~matches.

Before introducing the model, it is important to highlight some key characteristics of volleyball which needs to be taken into account when modelling the data. First, unlike other sports, the total number of points scored in a volleyball match by each team is not typically enough to determine which of the two teams effectively won the match. For example, consider the situation where a match ends with the home team winning a total of $3$ sets with a relatively small point difference with respect to the opponent in each set, while the away team wins $2$ sets with a substantially larger point gap in each set. Then, it is possible that the winning team is associated with a lower number of points across all sets compared with those scored by the losing team. 
Second, the total number of points assigned to the two opposing teams at the end of a match may vary depending on the number of sets played. If the total number of sets played is $3$ or $4$, the winning and losing team earns $3$ and $0$ points, respectively; if the total number of sets played is $5$, the winning and losing team are given $2$ and $1$ points,~respectively.

Finally, we note that, although the total number of points scored in a match could be potentially infinite, the total number of points scored in a single set may or may not have an upper bound. For example, when a team scores $25$ points in the first set and the opponent has scored $23$ or less, the match ends; however, if the opponent has scored $24$, the match continues until the point difference between the two teams is equal to $2$. Ideally, set specific statistics for each match would provide more detailed information about the performance of each team compared with using the aggregated points across all the sets played. However, this type of information is not typically available to the public and their analysis would also introduce additional modelling issues which are discussed in~Section~\ref{discussion}.

\section{Modelling Framework}\label{modelframe}
From a statistical perspective, one of the most interesting aspect in modelling team sport data is related to the distributional form associated with the number of scores in a single game by the two opponent teams. In the analysis of football data, the Poisson distribution has been widely used to model the game scores; for example, \cite{Karlis2003} proposed the use of a bivariate Poisson distribution which includes a parameter that explicitly accounts for the covariance between the goals scored by the two competing teams. More recently, \cite{Baio2010} adopted a Bayesian hierarchical model which simultaneously accounts for the correlation between the two variables via the specification of two conditionally independent Poisson variables for the number of goals scored. Being defined within a Bayesian framework, this flexible modelling specification also allows to generate predictions of future games in a relatively easy way by means of the posterior predictive distribution. 

We extend and adapt the modelling frameworks typically used for the analysis of football data and propose a novel Bayesian hierarchical modelling framework for the analysis and prediction of volleyball results in regular seasons. Three different sub-models or "modules" form our framework: (1) The module of the observed number of points scored by the two opposing teams in a match ($y_h$ and $y_a$); (2) the module of the binary indicator for the number of sets played ($d^s$); (3) the module of the binary indicator for the winner of the match ($d^m$). These three modules are jointly modelled using a flexible Bayesian parametric approach, which allows to fully propagate the uncertainty for each unobserved quantity and to assess the predictive performance of the model in a relatively easy way. In the following, we describe the notation and the model used in each of the three modules. 

\subsection{Module 1: Modelling the Scoring Intensity}\label{module1}
In the first module of the framework, we follow \cite{Karlis2003, Baio2010} and model the number of points scored by the home and away team in the $i$-th match of the season $\bm y=(y_{hi},y_{ai})$ using two independent Poisson distributions 
\begin{equation}\label{eqpois}
\begin{aligned}
y_{hi} &\sim \mbox{Poisson}(\theta_{hi})\\
y_{ai} &\sim \mbox{Poisson}(\theta_{ai}),
\end{aligned}
\end{equation}
conditionally on the set of parameters $\bm \theta=(\theta_{hi},\theta_{ai})$, representing the scoring intensity in the $i$-th match for the home and away team, respectively. The parameters in Equation~\ref{eqpois} are then modelled using the log-linear regressions
\begin{equation}\label{eqlog}
\begin{aligned}
\log(\theta_{hi}) &=\mu + \lambda + att_{h(i)} + def_{a(i)}\\
\log(\theta_{ai}) &=\mu + att_{a(i)} + def_{h(i)},
\end{aligned}
\end{equation}
which corresponds to a Poisson log-Normal model, whose parameters can be given specific interpretations~\citep{Karlis2003}. Within the formulae of Equation~\ref{eqlog}, $\mu$ is a constant, while $\lambda$ can be identified as the home effect and represents the advantage for the team hosting the game which is typically assumed to be constant for all the teams and throughout the season. The overall offensive and defensive performances of the $k$-th team is captured by the parameters $att$ and $def$, whose nested indexes $h(i), a(i)=1,\ldots,K$ identify the home and away team in the $i$-th game of the season, where $K$ denotes the total number of the teams. 

We then expand the modelling framework to incorporate match-specific statistics related to the offensive and defensive performances of the home and away teams. More specifically, the effects associated with the attack intensity of the home teams and the defence effect of the away teams are:
\begin{equation}\label{eq1}
\begin{aligned}
att_{h(i)} &=\alpha_{0h(i)} + \alpha_{1h(i)}att^{eff}_{hi}+ \alpha_{2h(i)}ser^{eff}_{hi}\\
def_{a(i)} &=\beta_{0a(i)} + \beta_{1a(i)}def^{eff}_{ai}+ \beta_{2a(i)}blo^{eff}_{ai}.
\end{aligned}
\end{equation}
We omit the index $i$ from the terms to the left-hand side of the above formulae to ease notation, i.e.~$att_{h(i)}=att_{h(i)i}$ and $def_{a(i)}=def_{a(i)i}$. The overall offensive effect of the home teams in Equation~\ref{eq1} is a function of a baseline team-specific parameter $\alpha_{0h(i)}$, and the attack and serve efficiencies of the home team, whose impact is captured by the parameters $\alpha_{1h(i)}$ and $\alpha_{2h(i)}$. The overall defensive effect of the away team is a function of a baseline team-specific parameter $\beta_{0a(i)}$, and the defence and block efficiencies of the away team, whose impact is captured by the parameters $\beta_{1a(i)}$ and $\beta_{2a(i)}$, respectively. 

Similarly, the effects associated with the attack intensity of the away teams and the defence effect of the home teams are:
\begin{equation}\label{eq2}
\begin{aligned}
att_{a(i)} &=\alpha_{0a(i)} + \alpha_{1a(i)}att^{eff}_{ai}+ \alpha_{2a(i)}ser^{eff}_{ai}\\
def_{h(i)} &=\beta_{0h(i)} + \beta_{1h(i)}def^{eff}_{hi}+ \beta_{2h(i)}blo^{eff}_{hi},
\end{aligned}
\end{equation}
where the overall offensive effect in Equation~\ref{eq2} depends on the parameters $\alpha_{0a(i)}$, $\alpha_{1a(i)}$ and $\alpha_{2a(i)}$, while the overall defensive effect depends on the parameters $\beta_{0h(i)}$, $\beta_{1h(i)}$ and $\alpha_{2h(i)}$.

To achieve identifiability of the model, a set of parametric constraints needs to be imposed. Following~\cite{Karlis2003, Baio2010}, we impose \textit{sum-to-zero} constraints on the team-specific parameters
\begin{equation}\label{eqcons}
\sum_{k=1}^{K}\alpha_{jk}=0 \;\;\; \text{and} \;\;\; \sum_{k=1}^{K}\beta_{jk}=0,
\end{equation}
for $k=1,\ldots,K$ and $j=(0,1,2)$. Under this set of constraints, the overall offensive and defensive effects of the teams are expressed as departures from a team of average offensive and defensive performance. An alternative set of constraints that could be imposed to achieve identifiability is known as \textit{corner} constraints, where the overall offensive and defensive effects of a "reference" team are set to $0$, and the effects of all other teams are expressed as departures from the effects of the reference team. However, we find the interpretation of the model parameters under Equation~\ref{eqcons} to be more intuitive compared with the corner constraints, without the potential issue of the choice of the reference team.

Within a Bayesian framework, prior distributions need to be specified for all random parameters in the model. Weakly informative Normal distributions centred at $0$ with a relatively large variances are specified for the fixed effect parameters (note that we specify Normal distributions in terms of mean and precision, where precision=1/variance)
\begin{equation*}
\mu \sim \text{Normal}(0,0.000001) \;\;\; \text{and} \;\;\; \lambda \sim \text{Normal}(0,0.000001).
\end{equation*}
The team-specific random effect parameters are instead modelled as exchangeable from common distributions 
\begin{equation*}
\alpha_{jk} \sim \text{Normal}(\mu_{\alpha_{j}},\tau_{\alpha_{j}}) \;\;\; \text{and} \;\;\; \beta_{jk} \sim \text{Normal}(\mu_{\beta_{j}},\tau_{\beta_{j}}),
\end{equation*}
for $k=1,\ldots,K$ and $j=(0,1,2)$. Finally, the mean and precision of the team-specific effect distributions are modelled using weakly informative Normal and Gamma priors, respectively:
\begin{equation*}
\begin{aligned}
\mu_{\alpha_{j}}& \sim \text{Normal}(0,0.000001), \;\;\; \mu_{\beta_{j}}  \sim \text{Normal}(0,0.000001), \\
\tau_{\alpha_{j}}& \sim \text{Gamma}(0.01,0.01), \;\;\; \tau_{\beta_{j}} \sim \text{Gamma}(0.01,0.01),
\end{aligned}
\end{equation*}
for $j=(0,1,2)$. These hyperparameters represent the latent structure of the model which affects the estimation of the scoring intensity parameters $\bm \theta$, and thus indirectly takes into account a form of correlation between the two outcome variables $y_{hi}$ and~$y_{ai}$. 

\subsection{Module 2: Modelling the Probability of Playing 5 Sets}\label{module2}
In the second module, we explicitly model the chance of playing $5$ sets in the $i$-th match of the season, i.e.~the sum of the sets won by the home ($s_{hi}$) and away ($s_{ai}$) team is equal to $5$. This is necessary when generating predictions in order to correctly assign the points to the winning/losing teams throughout the season and evaluate the rankings of the teams at the end of the season (Section~\ref{data}).

We model the indicator variable $d^s_{i}$, taking value $1$ if $5$ sets were played in the $i-$th match and $0$ otherwise, using a Bernoulli distribution
\begin{equation} \label{linpred_sets}
\begin{aligned}
d^s_{i}:&=\mathbb{I}(s_{hi}+s_{ai}=5)\sim\mbox{Bernoulli}(\pi^s_{i}) \\
\mbox{logit}(\pi^s_{i})&= \gamma_0 + \gamma_1y_{hi}+\gamma_2y_{ai}  
\end{aligned}
\end{equation}
where $\pi^s_{i}$ is the match probability of playing $5$ sets, which is estimated on the logit scale as a function of a set of parameters $\bm \gamma$. These include a baseline parameter $\gamma_0$ and the total number of points scored by the home and away teams, whose effect is captured by the parameters $\gamma_1$ and $\gamma_2$, respectively.

\subsection{Module 3: Modelling the Probability of Winning the Match}\label{module3}
The last module deals with the chance of the home team to win the $i$-th match, i.e.~the total number of sets won by the home team  ($s_{hi}$) is larger than that of the away team ($s_{ai}$) -- we note that we could have also equivalently decided to model the chance of the away team to win the $i$-th match. This part of the model is again necessary when predicting the results for future matches, since the team associated with the higher number of points scored in the $i$-th match may not correspond to the winning team (Section~\ref{data}). 

We model the indicator variable $d^m_{i}$, taking value $1$ if the home team won the $i-$th match and $0$ otherwise, using another Bernoulli distribution
\begin{equation} \label{linpred_matches}
\begin{aligned}
d^m_{i}:&=\mathbb{I}(s_{hi}>s_{ai}) \sim\mbox{Bernoulli}(\pi^m_{i}) \\
\mbox{logit}(\pi^m_{i})&= \eta_0 + \eta_1y_{hi}+\eta_2y_{ai} + \eta_3 d^s_i
\end{aligned}
\end{equation}
where $\pi^m_{i}$ is the match probability for the home team to win the match, which is estimated on the logit scale as a function of a set of parameters $\bm \eta$. These include a baseline parameter $\eta_0$, the total number of points scored by the home and away teams, and the binary indicator associated with playing $5$ sets. The effects of the last three variables is captured by the parameters $\eta_1$, $\eta_2$ and $\eta_3$, respectively.

Figure~\ref{fig1} shows a graphical representation of the structure of the model as well as the contribution of each of the three modules to the framework.
\begin{center}
FIGURE 1 HERE
\end{center}
The framework corresponds to a joint distribution for all the observed quantities which are explicitly modelled. This is factored into the product of the marginal distribution of the total number of points scored by the two teams in each match, Module~1 -- $p(\bm y)$, the conditional distribution of the probability of playing $5$ sets in a match given $\bm y$, Module~2 -- $p(d^s_i \mid \bm y)$, and the conditional probability of winning the match given $\bm y$ and $d^s_i$, Module~3 -- $p(d^m_i\mid \bm y, d^s_i)$. Module~1 also includes the different in-game statistics as covariates in the model. These are related to the either the offensive (serve and attack efficiency) or defensive (defence and block efficiency) effects of the home and away teams in each match of the season, and are respectively denoted in the graph as $\bm x^{att}_{ti}=(ser^{eff}_{ti}, att^{eff}_{ti})$ and $\bm x^{def}_{ti}=(def^{eff}_{ti}, blo^{eff}_{ti})$ to ease notation, for $t=(h,a)$.

\section{Accounting for the multilevel correlation}\label{modelcorr}
Although the individual-level correlation between the observable variables $y_{hi}$ and $y_{ai}$ is taken into account through the hierarchical structure of the framework, a potential limitation of the model is that it ignores the possible multilevel correlation between the team-specific offensive $\alpha_{jk}$ and defensive $\beta_{jk}$ coefficients, for $j=(0,1,2)$ and $k=1,\ldots,K$. To show the implications of this assumption, it is convenient to represent the two categories of multilevel parameters, described in Section~\ref{module1}, using the matrix notation
\begin{equation}\label{mult}
\bm \alpha \sim \text{Normal}\left( \bm M_{\bm \alpha}, \bm \Sigma^{-1}_{\bm \alpha} \right) \;\;\; \text{and} \;\;\; \bm \beta \sim \text{Normal}\left( \bm M_{\bm \beta}, \bm \Sigma^{-1}_{\bm \beta} \right),
\end{equation}
where $\bm \alpha$ and $\bm \beta$ are the two $K\times 3$ matrices containing the offensive and defensive regression coefficients associated with each team in the league, respectively. The hyperparameters
\begin{equation*}
\bm M_{\bm \alpha}=\left( \mu_{\alpha_{0}},\mu_{\alpha_{1}},\mu_{\alpha_{2}} \right) \;\;\; \text{and} \;\;\; \bm M_{\bm \beta}=\left( \mu_{\beta_{0}},\mu_{\beta_{1}},\mu_{\beta_{2}} \right),
\end{equation*}
are two vectors of length $3$, representing the mean of the joint distributions of $\bm \alpha$ and $\bm \beta$, while 
\scriptsize
\begin{equation*}
 \bm \Sigma_{\bm \alpha}=\left( {\begin{array}{*{20}c}
   \sigma^{2}_{\alpha_0} & \rho_{0\alpha}\sigma_{\alpha_0}\sigma_{\alpha_1} & \rho_{1\alpha}\sigma_{\alpha_0}\sigma_{\alpha_2}  \\
    \rho_{0\alpha}\sigma_{\alpha_1}\sigma_{\alpha_0} & \sigma^{2}_{\alpha_1} &  \rho_{2\alpha}\sigma_{\alpha_1}\sigma_{\alpha_2}  \\  
     \rho_{1\alpha}\sigma_{\alpha_2}\sigma_{\alpha_0} &   \rho_{2\alpha}\sigma_{\alpha_1}\sigma_{\alpha_2} &  \sigma^{2}_{\alpha_2}\\  
 \end{array} } \right) \;\;\; \text{\normalsize and} \;\;\; 
 \bm \Sigma_{\bm \beta}=\left( {\begin{array}{*{20}c}
   \sigma^{2}_{\beta_0} & \rho_{0\beta}\sigma_{\beta_0}\sigma_{\beta_1} & \rho_{1\beta}\sigma_{\beta_0}\sigma_{\beta_2}  \\
    \rho_{0\beta}\sigma_{\beta_1}\sigma_{\beta_0} & \sigma^{2}_{\beta_1} &  \rho_{2\beta}\sigma_{\beta_1}\sigma_{\beta_2}  \\  
     \rho_{1\beta}\sigma_{\beta_2}\sigma_{\beta_0} &   \rho_{2\beta}\sigma_{\beta_1}\sigma_{\beta_2} &  \sigma^{2}_{\beta_2}\\  
 \end{array} } \right) ,
\end{equation*}
\normalsize
are two $3\times3$ matrices, whose elements form the covariance matrices of the distributions of $\bm \alpha$ and~$\bm \beta$ in Equation~\ref{mult}. In Section~\ref{module1}, we implicitly assumed independence between the multilevel parameters and specify prior distributions only on $\tau_{\alpha_j}=\frac{1}{\sigma^2_{\alpha_j}}$ and $\tau_{\beta_j}=\frac{1}{\sigma^2_{\beta_j}}$, i.e.~we set $\bm \rho_j=(\rho_{j\alpha},\rho_{j\beta})=\bm 0$ for $j=(0,1,2)$. 

To account for the possible multilevel correlation, prior distributions on $\bm \rho_j$ can be specified. One possibility is to directly specify separate priors on each component of the covariance matrices $\bm \Sigma$, including the correlation parameters. However, when the number of multilevel parameters is more than $2$, modelling $\bm \rho_j$ is challenging, since they all need to be constrained between $-1$ and $1$ while also ensuring the positive definiteness of the covariance matrices. An alternative approach is to specify Inverse-Wishart prior distributions on $\bm \Sigma$ (or equivalently Wishart priors on $\bm \Sigma^{-1}$). This is defined as $\text{Inv-Wishart}(\nu,\bm \Omega)$ and has the computational advantage of being the conjugate prior for the covariance matrix of a multivariate Normal distribution, but requires the specification of two hyperprior parameters. The scale matrix $\bm \Omega$, which must be positive definite with dimension equal to the covariance matrix, and the degrees of freedom $\nu$, which is bounded from below by the dimension of $\bm \Omega$. A general problem when setting these parameters is that, depending on the prior choice, some undesirable constraints may be imposed on some of the parameters of~$\bm \Sigma$. For example, setting $\nu$ equal to the dimension of $\bm \Omega+1$ corresponds to specifying uniform priors on the correlation parameters $\bm \rho_j$, but it may lead to too restrictive priors on the scale parameters of~$\bm \Sigma$. Conversely, increasing the value of $\nu$ allows to estimate the scale parameters of $\bm \Sigma$ with more freedom, but at the cost of constraining the correlation parameters~\citep{Gelmanb}.

In our analysis, we account for the multilevel correlation using Inverse-Wishart distributions on $ \bm \Sigma_{\bm \alpha}$ and $ \bm \Sigma_{\bm \beta}$, which are scaled in order to facilitate the specification of the priors. This prior is known as the scaled Inverse-Wishart distribution~\citep{Gelmanb} and can be constructed in the following way. First, two vectors of scale parameters $\bm \xi_{\bm \alpha}$ and $\bm \xi_{\bm \beta}$ of length $3$ are introduced such that the covariance matrices can be decomposed as the products
\begin{equation}\label{invprior}
\bm \Sigma_{\bm \alpha}=\text{Diag}(\bm \xi_{\bm \alpha})\bm \Lambda_{\bm \alpha}\text{Diag}(\bm \xi_{\bm \alpha}) \;\;\; \text{and} \;\;\; \bm \Sigma_{\bm \beta}=\text{Diag}(\bm \xi_{\bm \beta}) \bm \Lambda_{\bm \beta}\text{Diag}(\bm \xi_{\bm \beta}),
\end{equation}
where $\text{Diag}(\bm \xi_{\bm \alpha})$ and $\text{Diag}(\bm \xi_{\bm \beta})$ are two $3\times3$ diagonal matrices with the entries of $\bm \xi_{\bm \alpha}$ and $\bm \xi_{\bm \beta}$ being the diagonal elements. The Inverse-Wishart priors are then specified on the unscaled covariance matrices $\bm \Lambda_{\bm \alpha}$ and $\bm \Lambda_{\bm \beta}$, that is
\begin{equation*}
\bm \Lambda_{\bm \alpha} \sim \text{Inv-Wishart}(\nu_{\bm \alpha},\bm \Omega_{\bm \alpha}) \;\;\; \text{and} \;\;\; \bm \Lambda_{\bm \beta} \sim \text{Inv-Wishart}(\nu_{\bm \beta},\bm \Omega_{\bm \beta}).
\end{equation*}
We set $\bm \Omega_{\bm \alpha}$ and $\bm \Omega_{\bm \beta}$ as identity matrices, and set $\nu_{\bm \alpha}$ and $\nu_{\bm \beta}$ equal to $4$. Weakly informative priors are then assigned to each scaling factor, i.e.~$\xi_{\alpha_j} \sim \text{Normal}(0,100)$ and $\xi_{\beta_j} \sim \text{Normal}(0,100)$, for~$j=(0,1,2)$.
Under this approach, the $j$-th variance components $\sigma^{2}_{\alpha_j}$ and $\sigma^{2}_{\beta_j}$ of the covariance matrices can then be retrieved as the product between the corresponding diagonal elements of $\bm \Lambda_{\bm \alpha}$ and $\bm \Lambda_{\bm \beta}$ and the appropriate scaling factors from $\bm \xi_{\bm \alpha}$ and $ \bm\xi_{\bm \beta}$:
\begin{equation*}
\sigma^{2}_{\alpha_j}= \Sigma_{ \alpha_{jj}}=\xi^2_{ \alpha_{j}} \Lambda_{ \alpha_{jj}} \;\;\; \text{and} \;\;\; \sigma^{2}_{\beta_j}=\Sigma_{ \beta_{jj}}=\xi^2_{ \beta_{j}} \Lambda_{ \beta_{jj}},
\end{equation*}
for $j=(0,1,2)$. Similarly, the $jl$-th covariance components are given by
\begin{equation*}
\Sigma_{ \alpha_{jl}}=\xi_{ \alpha_{j}}\xi_{ \alpha_{l}} \Lambda_{ \alpha_{jl}}\;\;\; \text{and} \;\;\; \Sigma_{ \beta_{jl}}=\xi_{\beta_{j}}\xi_{\beta_{l}} \Lambda_{\beta_{jl}},
\end{equation*}
for $j,l=(0,1,2)$ and $j\neq l$. Since this parametrisation is based on the decomposition of the covariance matrix $\bm \Sigma$, the parameters $\bm \xi$ and $\bm \Lambda$ cannot be interpreted separately. However, Equation~\ref{invprior} provides a useful way to set up the model and minimise the amount of undesirable prior constraints when estimating the parameters of interest.

\section{Implementation and Results}\label{results}

We fitted all models using \texttt{JAGS}~\citep{Plummer}, a software specifically designed for the analysis of Bayesian models using Markov Chain Monte Carlo (MCMC) simulation~\citep{Brooks}, which was interfaced with \texttt{R} through the package \texttt{R2jags}~\citep{Su}. Samples from the posterior distribution of the parameters of interest generated by \texttt{JAGS} and saved to the \texttt{R} workspace were then used to produce summary statistics and plots. We ran two chains with 20,000 iterations per chain, using a burn-in of 10,000, for a total sample of 20,000 iterations for posterior inference. For each unknown quantity in the model, we assessed convergence and autocorrelation of the MCMC simulations using diagnostic measures such as the \textit{potential scale reduction factor} and the \textit{effective sample size}~\citep{Gelman2}. Alternative prior distributions were considered to check that no unintended information was included into the models through the priors. For example, weakly informative uniform priors for the standard deviations and different values for the variance of normally distributed regression parameters were considered. Results were robust to these specifications. We provide in Appendix~\ref{A1} the \texttt{JAGS} code to implement the models.

\subsection{Results}\label{res}
In this section, we present the results of the models from a twofold perspective. First, we look at the estimates of the team-specific marginal offensive and defensive effects. This is achieved by setting all the attack and defence related covariates in Equation~\ref{eq1} and Equation~\ref{eq2} to their centered version, i.e.~subtracting the means from the original vectors. In this way, it is possible to interpret the intercept parameters $\alpha_{0k}$ and $\beta_{0k}$ as the marginal offensive and defensive effects associated with each team in the league. Second, we assess and compare the ability of the models to predict future results. Specifically, we use the estimates of $\bm \theta$ at each posterior iteration to generate a vector $\bm y^{pred}$ of $1000$ replications from the posterior predictive distribution of $\bm y$. This, in turn, is used to validate the model and to predict the results of matches from a future (exchangeable) season.

\subsubsection{Posterior Estimates}\label{estpost}
Figure~\ref{fig11} displays the average marginal mean offensive and defensive effects for each of the $12$ teams in the league, which are estimated using the basic model of Section~\ref{modelframe} and the scaled Inverse-Wishart (IW) model of Section~\ref{modelcorr}. 
\begin{center}
FIGURE 2 HERE
\end{center}
The position of the points in the graphs is related to the average offensive and defensive performance of the teams, which are in turn based on the number of points scored and conceded by each team throughout the season. We note that, the teams who perform better are those associated with positive attack effects (more points scored) and negative defence effects (less points conceded).

The results from both models are similar with limited differences. In general, we can identify four main clusters of teams. The first cluster is formed by a single team (\textit{Novara}) who is associated with the largest attack effect among all the teams, but with a close-to-zero defence effect (the effect is estimated to be slightly negative and positive under the first and second model, respectively). The second cluster is formed by two teams (\textit{Conegliano} and \textit{Scandicci}) who have small positive attack effects but relatively large negative defence effects; \textit{Scandicci} is by far the team with the smallest defence effect (reflecting the fact that it has the minimum number of conceded points in the league).The third cluster is formed by six teams whose offensive and defensive effects place them in the top part of the graphs; among these, \textit{Casalmaggiore} is the team with the largest attack and smallest defence effect, while \textit{Bergamo} and \textit{San Casciano} are associated with negative attack effects. Finally, the fourth cluster is formed by three teams (\textit{Filottrano}, \textit{Legnano} and \textit{Pesaro}) who are located on the top-left part of the graphs, and are associated with the poorest offensive performances. Additional posterior summary measures, including the posterior 95\% credible intervals, for the team-specific attack and defence effects, the home effect ($\lambda$) and the constant ($\mu$) parameters are provided in Table~\ref{tab1app} and Table~\ref{tab2app} in Appendix~\ref{A2}.

\subsubsection{Predictions}\label{pred}
We proceed to assess the predictive performance of the models using the observed and replicated data. The latter are generated from the posterior predictive distributions of the models using the posterior samples of the parameters indexing the three modules of the framework: the scoring intensity rates $\bm \theta$ (Module 1), the probability of playing $5$ sets $\pi^s$ (Module 2), and the probability of winning the match $\pi^m$ (Module 3). We use the posterior estimates from these parameters to jointly sample $1000$ replications of the data $(\bm y^{rep},d^{s,rep},d^{m,rep})$, which are then used for model assessment. Table~\ref{tab2} compares the predictive performance of the models in terms of the total number of points scores and conceded, the number of matches won and the number of points in the league earned by each team at the end of the season. 
\begin{table}[!h]
\caption{Posterior predictive validation of the models. The observed data and the replications from the basic and the scaled IW model are compared in terms of: total number of points scores and conceded, total number of wins and total number of points earned by each team at the end of the season.}\label{tab2}
\centering
\scalebox{0.73}{
\begin{tabular}{ccccccccccccccc}
  \toprule
  & \multicolumn{4}{c}{Observed} & &  \multicolumn{4}{c}{Basic model} & & \multicolumn{4}{c}{Scaled IW model}\\ \cline{2-5} \cline{7-10} \cline{12-15} \noalign{\smallskip} 
Teams & Scored & Conc'd & Wins & Points        &        & Scored & Conc'd & Wins & Points       &       & Scored & Conc'd & Wins & Points \\
  \midrule
Bergamo & 1848 & 2025 & 7 & 19         &        & 1846 & 2020 & 7 & 21        &       & 1847 &  2020 & 7 &21\\ 
  Busto Arsizio & 1999 & 1927 & 12 & 39  &      & 1998 & 1919 & 12 & 37      &       & 2006 & 1920 & 12 & 37\\ 
  Casalmaggiore & 1922 & 2051 & 6 & 23 &     & 1918 & 2042 & 7 & 23         &     & 1906& 2039 & 7 &23\\ 
  Conegliano & 1960 & 1696 & 17 & 50    &     & 1960 & 1706 & 18 & 50       &      & 1955& 1713 & 18 & 50\\ 
  Filottrano & 1781 & 1961 & 7 & 19       &       & 1790 & 1954 & 6 & 18         &     & 1810& 1950 & 6 & 18\\ 
  Legnano & 1642 & 1903 & 5 & 11         &      & 1656 & 1898 & 4 & 17         &      & 1660& 1901 &  4& 17\\ 
  Monza & 2003 & 1943 & 13 & 37           &     & 1994 & 1938 & 13 & 38        &        & 2005& 1935 &13 & 38\\ 
  Novara & 1987 & 1776 & 17 & 51         &     & 1963 & 1776 & 17 & 51       &        & 1963& 1781 & 17& 51\\ 
  Pesaro & 1776 & 1820 & 10 & 32          &     & 1785 & 1828 & 11 & 32        &        & 1789& 1831 & 11 & 33\\ 
  Piacenza & 1888 & 1939 & 12 & 33       &     & 1886 & 1935 & 9 & 30           &     & 1890& 1933 & 9 & 30\\ 
  San Casciano & 1807 & 1881 & 8 & 27  &     & 1812 & 1880 & 9 & 29           &     & 1797& 1879 & 9 & 28\\ 
  Scandicci & 1865 & 1556 & 18 & 50       &     & 1858 & 1578 & 18 & 51         &     & 1862&  1580 & 18 & 51\\ 
   \bottomrule
\end{tabular}
}
\end{table}
Overall, the predicted results from both the basic and the scaled IW model seem to replicate the observed data relatively well for most of the teams. The total number of points scored and conceded are similar between the observed and replicated data, with the teams scoring (conceding) the most being also associated with the highest replicated points scored (conceded) and vice versa. Relatively small discrepancies are observed between the results of the two models for some of the teams. The total number of wins and league points are almost identical between the observed and replicated data, with the scaled IW model being associated with slightly more accurate predictions compared with the basic model. 

Figure~\ref{fig2} compares the cumulative points derived from the observed results throughout the season (the black line) and the predictions from both the basic model (in red), and the scaled Inverse-Wishart model (in blue).
\begin{center}
FIGURE 3 HERE
\end{center}
For almost all teams the predicted results are relatively close to the observed data and suggest a good performance of both models. In particular, the lines of the observed and replicated cumulative points intertwine in many cases, with only two teams for which the results are slightly overestimated (\textit{Filottrano} and \textit{San Casciano}). We note that, for some of the teams, the predictions from the scaled IW model are closer to the line of the observed points and outperform the predictions from the basic model (\textit{Bergamo}, \textit{Busto Arsizio}, \textit{Conegliano}, \textit{Monza}, \textit{San Casciano} and \textit{Scandicci}). For all the other teams, the performance of the two models is either similar or varies throughout the~season.

Finally, we compare the predictive ability of the two models in terms of the rankings of the teams in the league (calculated based on the total number of points earned by each team at the end of the season). Figure~\ref{fig3} shows the probability of ending the season in different positions for each team in the league, which is calculated based on the replications from either the basic (top graph) or scaled IW (bottom graph) model. Each position is associated with a different colour, from dark blue ($1$-st position) to yellow ($12$-th position).
\begin{center}
FIGURE 4 HERE
\end{center}
The predictive probabilities from both models are very similar and are in line with the observed rankings of the teams in the league (see Table~\ref{tab2}). There are some variations (from $1\%$ to $7\%$) in the probability of being in a specific rank between the results from the two models for some of the teams, but in most of the cases these differences remain negligible.

\section{Discussion}\label{discussion}
The model presented in this paper is an application of Bayesian hierarchical modelling to sport data. The basic structure presented in Section~\ref{modelframe} and its extension in Section~\ref{modelcorr} can be easily implemented and run using standard MCMC
algorithms, such as the one provided for \texttt{JAGS} in Appendix~\ref{A1}. Previous approaches have focused either on the identification of the key factors determining the performance of a team~\citep{Gaetan,Zhang2016} or the prediction of team rankings~\citep{Tumer2017}. To our knowledge, this is the first modelling framework which jointly allows to predict team rankings and the outcomes of the matches during a season in volleyball. The two alternative specifications implemented in our analysis show generally good predictive performances; between the two models, the scaled IW model seems to be slightly more accurate compared with the basic model, but is also associated with a higher level of complexity.

One potential limitation of the framework is that only match-specific statistics are used for estimation and prediction purposes. Ideally, the use of set-specific statistics could improve the predictive power of the model while also eliminating the need to include the models for the indicator variables of Equation~\ref{linpred_sets} and Equation~\ref{linpred_matches} . Indeed, since predictions would be related to the results from each set played, the information about the number of sets and the winner of the match would be directly incorporated. However, this would introduce additional problems related to the choice of the distributions for modelling the number of points scored by the opposing teams in a set, which is subject to specific constraints (e.g.~difference of $2$ points if one team scores more than $25$). In addition, set-specific statistics are not typically available from national federation web sites and are difficult to collect for all the matches in a season. 

Finally, the flexibility of the proposed framework allows the extension of the model in many ways. For example, additional types of in-game statistics (e.g.~number of passes), if available, could be incorporated to further improve the predictions of the model; alternative distributions could also be specified to model the total number of scores $\bm y$ during the season (e.g.~Negative Binomial). The scaled IW model could be extended by jointly modelling the team-specific  attack and defence effects ($\bm \alpha, \bm \beta$) with a single multivariate Normal distribution and scaled IW prior to account for the possible multilevel correlation between all random effect parameters. Although in our analysis the implementation of this more complex model did not lead to any substantial variation in the results, compared with the model of Section~\ref{modelcorr}, this may not be the case in other applications. In addition, following the approaches in the recent literature on the analysis of football matches~\citep{Tsokos2019}, data from different leagues and seasons could also be analysed jointly to deliver a context-specific validation framework that accounts for the temporal dimension in the data and the clustering across the leagues.

\bibliographystyle{tfs}
\bibliography{volleyball}


\appendix
\renewcommand\thefigure{\arabic{figure}}   

\section{Model Code}\label{A1}
The complete \texttt{JAGS} code for the models in Section~\ref{modelframe} and Section~\ref{modelcorr} used in the analysis is given~below.
\begin{Verbatim}[tabsize=8,fontsize=\footnotesize]
model {

# LIKELIHOOD FOR THE THREE MODULES

for (i in 1:nmatches) {

# Observed number of points scored by each team (Module 1)
y.h[n] ~ dpois(theta[i,1])
y.a[n] ~ dpois(theta[i,2])

##scoring intensities (accounting for mixing components)
log(theta[i,1]) <- mu + lambda + att[i,1] + def[i,1]
att.h[i,1] <- alpha0[hometeam[i]] + alpha1[hometeam[i]]*att.eff.h[g] + 
                     alpha2[hometeam[i]]*ser.eff.h[i] 
def.h[i,1] <- beta0[awayteam[i]] + beta1[awayteam[i]]*def.eff.a[g] +
                     beta2[awayteam[i]]*blo.eff.a[i]

log(theta[i,2]) <- mu + att.a[i,2] + def.a[i,2]
att.a[i,2] <- alpha0[awayteam[i]] + alpha1[awayteam[i]]*att.eff.a[i] + 
                     alpha2[awayteam[i]]*ser.eff.a[i] 
def.a[i,2] <- beta0[hometeam[i]] + beta1[hometeam[i]]*def.eff.h[i] + 
                     beta2[hometeam[i]]*blo.eff.h[i]


# Indicators for number of sets played (Module 2)
d.s[i] ~ dbern(pi.s[i])
logit(pi.s[i]) <- gamma[1] + gamma[2]*y.h[i] + gamma[3]*y.a[i]


# Indicators for match winner (Module 3)
d.m[i] ~ dbern(pi.m[i])
logit(pi.m[i]) <- eta[1] + eta[2]*y.h[i] + eta[3]*y.a[i] + eta[4]*d.s[i]
}

# Priors on the constant and home effects
mu ~ dnorm(0,0.00001)
home ~ dnorm(0,0.00001)

########################

# INDEPENDENT PRIORS: BASIC MODEL

## Trick to code the ‘‘sum-to-zero’’ constraint
for (t in 1:nteams){
alpha0.star[t] ~ dnorm(mu.alpha0,tau.alpha0)
beta0.star[t] ~ dnorm(mu.beta0,tau.beta0)
alpha0[t] <- alpha0.star[t] - mean(alpha0.star[])
beta0[t] <- beta0.star[t] - mean(beta0.star[])
alpha1.star[t] ~ dnorm(mu.alpha1,tau.alpha1)
beta1.star[t] ~ dnorm(mu.beta1,tau.beta1)
alpha2.star[t] ~ dnorm(mu.alpha2,tau.alpha2)
beta2.star[t] ~ dnorm(mu.beta2,tau.beta2)
alpha1[t] <- alpha1.star[t] - mean(alpha1.star[])
beta1[t] <- beta1.star[t] - mean(beta1.star[])
alpha2[t] <- alpha2.star[t] - mean(alpha2.star[])
beta2[t] <- beta2.star[t] - mean(beta2.star[])
}

##priors on the random effects
mu.alpha0 ~ dnorm(0,0.00001)
mu.beta0 ~ dnorm(0,0.00001)
tau.alpha0 ~ dgamma(0.01,0.01)
tau.beta0 ~ dgamma(0.01,0.01)
mu.alpha1 ~ dnorm(0,0.00001)
mu.beta1 ~ dnorm(0,0.00001)
tau.alpha1 ~ dgamma(0.01,0.01)
tau.beta1 ~ dgamma(0.01,0.01)
mu.alpha2 ~ dnorm(0,0.00001)
mu.beta2 ~ dnorm(0,0.00001)
tau.alpha2 ~ dgamma(0.01,0.01)
tau.beta2 ~ dgamma(0.01,0.01)

# JOINT PRIOR: SCALED IW MODEL 

## Trick to code the ‘‘sum-to-zero’’ constraint
for (t in 1:nteams){
alpha0.star[t] <- xi.alpha0*Alpha[t,1]
alpha1.star[t] <- xi.alpha1*Alpha[t,2]
alpha2.star[t] <- xi.alpha2*Alpha[t,3]
beta0.star[t] <- xi.beta0*Beta[t,1]
beta1.star[t] <- xi.beta1*Beta[t,2]
beta2.star[t] <- xi.beta2*Beta[t,3]

alpha0[t] <- alpha0.star[t] - mean(alpha0.star[])
beta0[t] <- beta0.star[t] - mean(beta0.star[])
alpha1[t] <- alpha1.star[t] - mean(alpha1.star[])
beta1[t] <- beta1.star[t] - mean(beta1.star[])
alpha2[t] <- alpha2.star[t] - mean(alpha2.star[])
beta2[t] <- beta2.star[t] - mean(beta2.star[])

#multivariate normal prior on random effects
Alpha[t,1:3] ~ dmnorm (M.raw.alpha[t,], Tau.raw.alpha[,])
Beta[t,1:3] ~ dmnorm (M.raw.beta[t,], Tau.raw.beta[,])

#raw prior mean
M.raw.alpha[t,1] <- mu.raw.alpha0
M.raw.alpha[t,2] <- mu.raw.alpha1
M.raw.alpha[t,3] <- mu.raw.alpha2
M.raw.beta[t,1] <- mu.raw.beta0
M.raw.beta[t,2] <- mu.raw.beta1
M.raw.beta[t,3] <- mu.raw.beta2
}

#raw hyperprior means
mu.raw.alpha0 ~ dnorm (0, 0.00001)
mu.raw.alpha1 ~ dnorm (0, 0.00001)
mu.raw.alpha2 ~ dnorm (0, 0.00001)
mu.raw.beta0 ~ dnorm (0, 0.00001)
mu.raw.beta1 ~ dnorm (0, 0.00001)
mu.raw.beta1 ~ dnorm (0, 0.00001)

#priors on the scaling factors
xi.alpha0 ~ dunif (0, 100)
xi.alpha1 ~ dunif (0, 100)
xi.alpha2 ~ dunif (0, 100)
xi.beta0 ~ dunif (0, 100)
xi.beta1 ~ dunif (0, 100)
xi.beta2 ~ dunif (0, 100)

#rescaled hyperprior means
mu.alpha0 <- xi.alpha0*mu.raw.alpha0
mu.alpha1 <- xi.alpha1*mu.raw.alpha1
mu.alpha2 <- xi.alpha2*mu.raw.alpha2
mu.beta0 <- xi.alpha0*mu.raw.beta0
mu.beta1 <- xi.alpha0*mu.raw.beta1
mu.beta2 <- xi.alpha0*mu.raw.beta2

#Wishart prior on raw precision matrix
Tau.raw.alpha[1:3,1:3] ~ dwish (Omega.alpha[,], nu.alpha)
Tau.raw.beta[1:3,1:3] ~ dwish (Omega.beta[,], nu.beta)

#raw covariance matrix
Sigma.raw.alpha[1:3,1:3] <- inverse(Tau.raw.alpha[,])
Sigma.raw.beta[1:3,1:3] <- inverse(Tau.raw.beta[,])

#rescaled standard deviation components
sigma.alpha0 <- xi.alpha0*sqrt(Sigma.raw.alpha[1,1])
sigma.alpha1. <- xi.alpha1*sqrt(Sigma.raw.alpha[2,2])
sigma.alpha2 <- xi.alpha2*sqrt(Sigma.raw.alpha[3,3])
sigma.beta0 <- xi.beta0*sqrt(Sigma.raw.beta[1,1])
sigma.beta1 <- xi.beta1*sqrt(Sigma.raw.beta[2,2])
sigma.beta2 <- xi.beta2*sqrt(Sigma.raw.beta[3,3])

##############################

#priors on the logistic regressions
for(s in 1:3){
gamma[s] ~ dnorm(0,0.0001)
}
for(k in 1:4){
eta[k] ~ dnorm(0,0.0001)
}

}
\end{Verbatim}

\section{Posterior Results}\label{A2}

\begin{table}[!h]
\caption{Posterior summaries of the estimates of the marginal team-specific attack and defence effects, the home effect and the constant from the log-linera regression of the basic model in Section~\ref{modelframe}. The model is fitted to data from the women's volleyball Italian Serie A1 season 2017-2018.}\label{tab1app}
\centering
\scalebox{0.8}{
\begin{tabular}{cccccccccccc}
  \toprule
            &   \multicolumn{5}{c}{Attack Effect} &  &\multicolumn{5}{c}{Defence Effect}\\ \cline{2-6} \cline{8-12} \noalign{\smallskip} 
 Teams & mean & sd & 2.5\% & median & 97.5\% & & mean & sd & 2.5\% &median & 97.5\% \\ 
  \midrule
Bergamo & -0.0098 & 0.0246 & -0.0589 & -0.0098 & 0.0382 & & 0.0434 & 0.0224 & -0.0005 & 0.0433 & 0.0873 \\ 
  Busto Arsizio & 0.0177 & 0.0223 & -0.0257 & 0.0178 & 0.0612 & & 0.0594 & 0.0276 & 0.0054 & 0.0594 & 0.1138 \\ 
  Casalmaggiore & 0.0247 & 0.0221 & -0.0190 & 0.0251 & 0.0676 & & 0.0859 & 0.0236 & 0.0399 & 0.0861 & 0.1320 \\ 
  Conegliano & 0.0042 & 0.0242 & -0.0433 & 0.0042 & 0.0518 & &-0.0846 & 0.0235 & -0.1307 & -0.0843 & -0.0389 \\ 
  Filottrano & -0.0750 & 0.0276 & -0.1299 & -0.0746 & -0.0219 & & 0.0318 & 0.0248 & -0.0172 & 0.0315 & 0.0799 \\ 
  Legnano & -0.0957 & 0.0260 & -0.1471 & -0.0957 & -0.0446 &  &0.0144 & 0.0242 & -0.0327 & 0.0143 & 0.0622 \\ 
  Monza & 0.0309 & 0.0239 & -0.0161 & 0.0307 & 0.0777 & &0.0491 & 0.0250 & -0.0001 & 0.0492 & 0.0976 \\ 
  Novara & 0.1448 & 0.0334 & 0.0807 & 0.1447 & 0.2103 & &-0.0069 & 0.0281 & -0.0612 & -0.0072 & 0.0487 \\ 
  Pesaro & -0.0527 & 0.0232 & -0.0982 & -0.0526 & -0.0078 & & -0.0326 & 0.0237 & -0.0792 & -0.0324 & 0.0131 \\ 
  Piacenza & 0.0105 & 0.0247 & -0.0375 & 0.0104 & 0.0594 & & 0.0311 & 0.0251 & -0.0178 & 0.0312 & 0.0800 \\ 
  San Casciano & -0.0140 & 0.0248 & -0.0628 & -0.0141 & 0.0345 & & 0.0087 & 0.0256 & -0.0423 & 0.0091 & 0.0582 \\ 
  Scandicci & 0.0144 & 0.0294 & -0.0428 & 0.0142 & 0.0727 & & -0.1998 & 0.0340 & -0.2680 & -0.1992 & -0.1351 \\ [1em]
  Home &  0.0343 & 0.0147 & 0.0054 & 0.0343 & 0.0633 & & & &  &  &  \\
  Constant & 4.443 & 0.012  & 4.419 &  4.443 &  4.467 & & & &  &  &\\
   \bottomrule
\end{tabular}
}
\end{table}

\begin{table}[!h]
\caption{Posterior summaries of the estimates of the marginal team-specific attack and defence effects, the home effect and the constant from the log-linear regression of the scaled IW model in Section~\ref{modelcorr}. The model is fitted to data from the women's volleyball Italian Serie A1 season 2017-2018.}\label{tab2app}
\centering
\scalebox{0.8}{
\begin{tabular}{cccccccccccc}
  \toprule
            &   \multicolumn{5}{c}{Attack Effect} &  &\multicolumn{5}{c}{Defence Effect}\\ \cline{2-6} \cline{8-12} \noalign{\smallskip} 
 Teams & mean & sd & 2.5\% & median & 97.5\% & & mean & sd & 2.5\% & median & 97.5\% \\ 
  \midrule
Bergamo & -0.0130 & 0.0211 & -0.0543 & -0.0131 & 0.0287 & & 0.0401 & 0.0224 & -0.0037 & 0.0401 & 0.0841 \\ 
  Busto Arsizio & 0.0188 & 0.0207 & -0.0224 & 0.0190 & 0.0589 & &0.0610 & 0.0289 & 0.0050 & 0.0610 & 0.1176 \\ 
  Casalmaggiore & 0.0138 & 0.0209 & -0.0269 & 0.0136 & 0.0555 & & 0.0817 & 0.0233 & 0.0364 & 0.0816 & 0.1272 \\ 
  Conegliano & 0.0011 & 0.0236 & -0.0452 & 0.0011 & 0.0473 & &-0.0792 & 0.0234 & -0.1256 & -0.0790 & -0.0341 \\ 
  Filottrano & -0.0573 & 0.0244 & -0.1086 & -0.0563 & -0.0128 & &0.0312 & 0.0242 & -0.0165 & 0.0313 & 0.0782 \\ 
  Legnano & -0.0854 & 0.0233 & -0.1329 & -0.0849 & -0.0412 & &0.0160 & 0.0234 & -0.0300 & 0.0159 & 0.0619 \\ 
  Monza & 0.0298 & 0.0206 & -0.0111 & 0.0300 & 0.0702 & &0.0424 & 0.0247 & -0.0054 & 0.0422 & 0.0913 \\ 
  Novara & 0.1530 & 0.0350 & 0.0850 & 0.1527 & 0.2221 & &0.0073 & 0.0303 & -0.0520 & 0.0070 & 0.0666 \\ 
  Pesaro & -0.0537 & 0.0207 & -0.0951 & -0.0535 & -0.0133 & &-0.0296 & 0.0238 & -0.0762 & -0.0297 & 0.0170 \\ 
  Piacenza & 0.0067 & 0.0220 & -0.0367 & 0.0069 & 0.0497 & &0.0309 & 0.0240 & -0.0163 & 0.0309 & 0.0781 \\ 
  San Casciano & -0.0291 & 0.0228 & -0.0725 & -0.0297 & 0.0175 & & 0.0068 & 0.0251 & -0.0434 & 0.0071 & 0.0554 \\ 
  Scandicci & 0.0155 & 0.0264 & -0.0355 & 0.0153 & 0.0686 & &-0.2084 & 0.0369 & -0.2818 & -0.2082 & -0.1360 \\ [1em]
  Home &  0.0311 & 0.0147 & 0.0022 & 0.031 & 0.0601 & & & &  &  &  \\
  Constant & 4.447 & 0.012  & 4.424 &  4.447 &  4.471 & & & &  &  &\\
   \bottomrule
\end{tabular}
}
\end{table}


\clearpage

\begin{figure}[!h]
\centering
\includegraphics[scale=1]{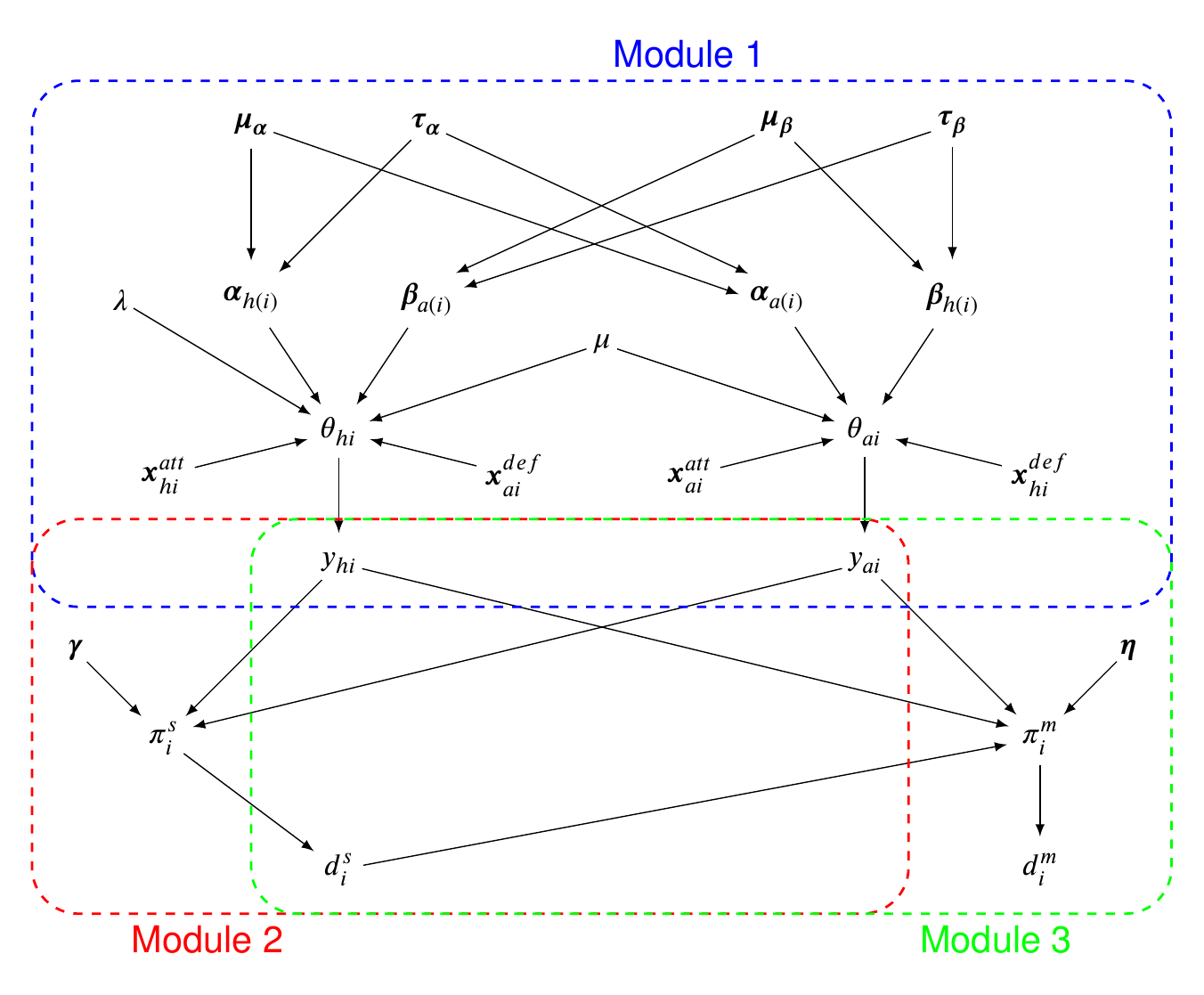}
\caption{Graphical representation of the modelling framework. The joint distribution $p(\bm y,d^s_i,d^m_i)$ is expressed in terms of a distribution of the scoring intensity $p(\bm y)$, the conditional distributions of the probability of playing $5$ sets $p(d^s_i \mid \bm y)$ and the conditional probability of winning the match $p(d^m_i\mid \bm y, d^s_i)$, respectively indicated with a dashed blue, red and green line. The parameters indexing the corresponding distributions or ``modules'' are indicated with different Greek letters, while $i$ denotes the match index within the regular season. The solid black arrows show the dependence relationships between the parameters and variables within and between the three modules.}\label{fig1}
\end{figure}

\begin{figure}[!h]
\centering
\subfloat[Basic Model]{\includegraphics[scale=0.5]{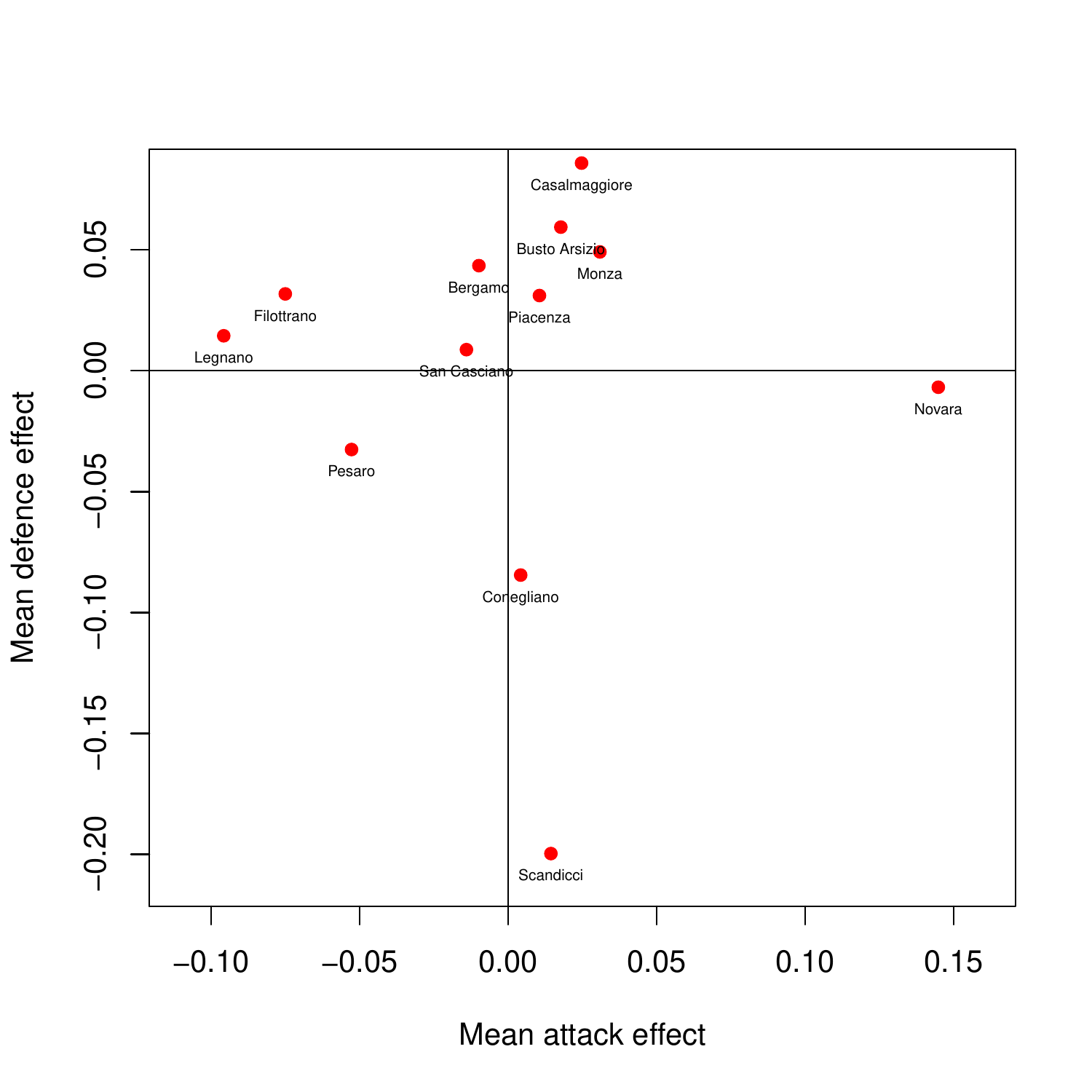}}
\subfloat[Scaled Inverse-Wishart Model]{\includegraphics[scale=0.5]{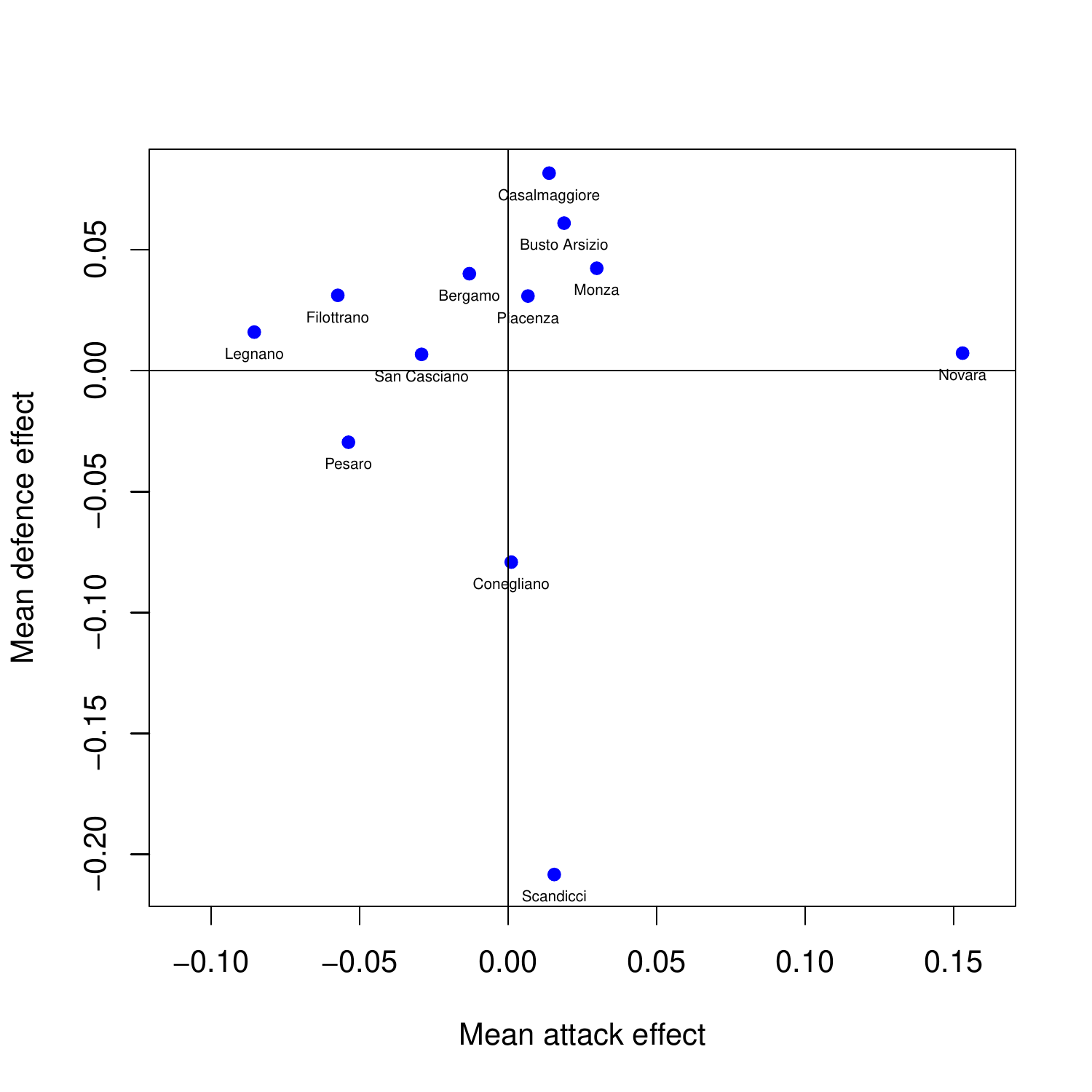}}
\caption{Average marginal mean offensive and defensive effects for all teams in the league, estimated from the log-linear regression of the basic model from Section~\ref{modelframe} (panel a) and the scaled IW model from Section~\ref{modelcorr} (panel b).}\label{fig11}
\end{figure}

\begin{figure}[!h]
\centering
\includegraphics[scale=0.7]{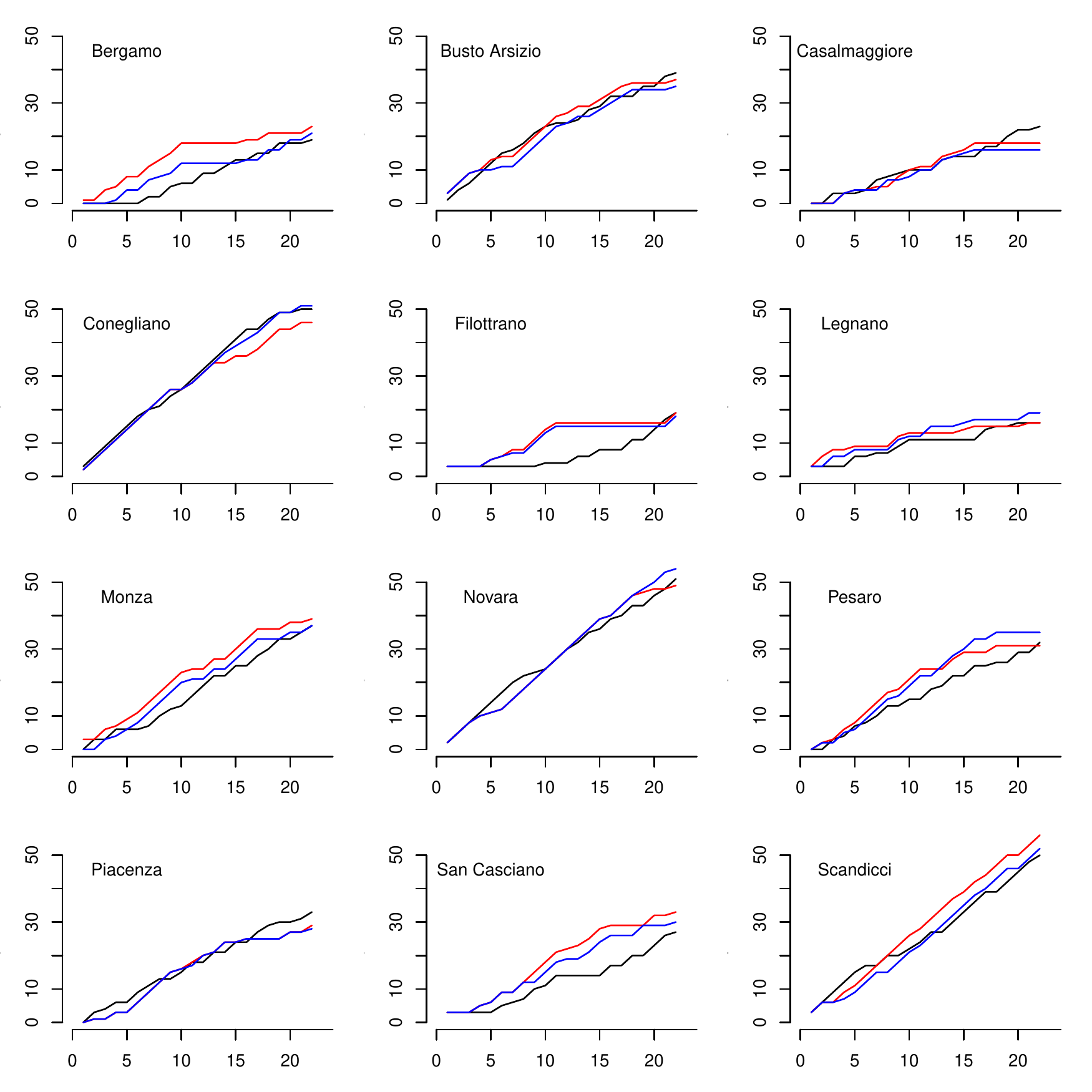}
\caption{Posterior predictive validation of the models. The black line represents the observed cumulative points through the season for each team in the league, while the red and blue lines represent the corresponding predictions from the basic and scaled IW model, respectively.}\label{fig2}
\end{figure}

\begin{figure}[!h]
\centering
\includegraphics[scale=0.85]{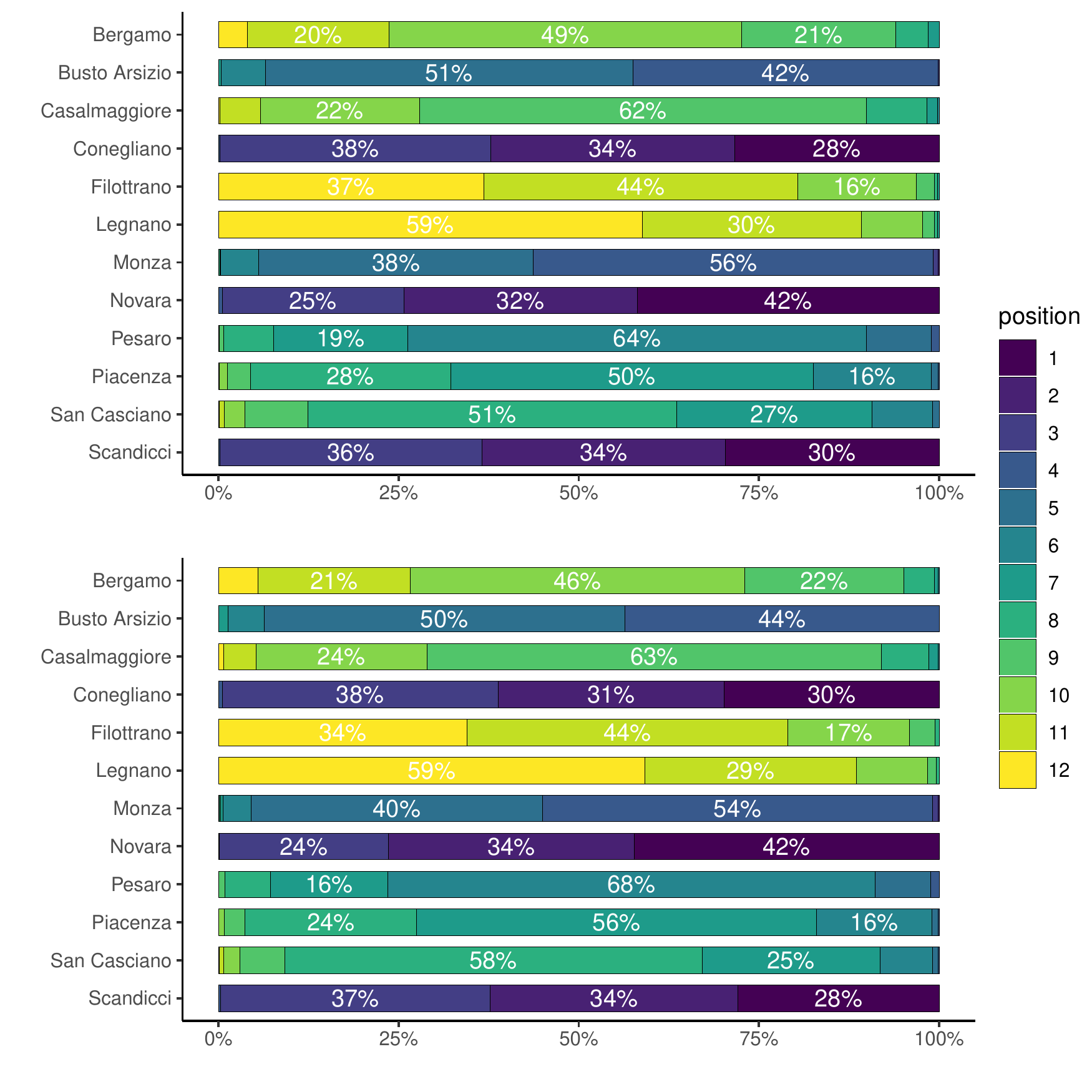}
\caption{Posterior probabilities of the rankings for each team in the league, estimated using the replicated data generated from the basic model (top panel) and the scaled IW model (bottom panel). Each position is associated with a different colour with higher (lower) ranks being associated with darker (lighter) colours. Only the probability values above 10\% are displayed for clarity.}\label{fig3}
\end{figure}

\end{document}